\documentclass[twocolumn,twoside,slac_two]{revtex4}
\usepackage{graphicx}
\usepackage{lscape}
\usepackage{rotating}
\usepackage{fancyhdr}
\begin{document}

\title{Earth Occultation Imaging Applied to BATSE -- Application to a Combined BATSE-GBM Survey of the Hard X-Ray Sky}

%

\author{Yuan Zhang, Michael Cherry}
\affiliation{Dept of Physics \& Astronomy, Louisiana State University, Baton Rouge, LA 70803}
\author{Gary Case}
\affiliation{Department of Physics, La Sierra University, Riverside, CA 92515}

\author{James Ling}
\affiliation{Jet Proporation Laboratory, NASA, Pasadena, CA 91109}

\author{William Wheaton}
\affiliation{Infrared Processing and Analysis Center, Caltech, Pasadena, CA 91125}

\begin{abstract}
A combined BATSE-GBM hard X-ray catalog is presented based on Earth Occultation Imaging applied to a reanalysis of BATSE data.
An imaging approach has been developed for the reanalysis of Earth Occultation analysis of BATSE data. The standard occultation analysis depends on a predetermined catalog of potential sources, so that a real source not present in the catalog may induce systematic errors when source counts associated with an uncatalogued source are incorrectly attributed to catalog sources. The goal of the imaging analysis is to find a complete set of hard X-ray sources, including sources not in the original BATSE occultation catalog. Using the imaging technique, we have identified 15 known sources and 17 unidentified sources and added them to the BATSE occultation catalog. The resulting expanded BATSE catalog of sources observed during 1991-2000 is compared to the ongoing GBM survey.

\end{abstract}

\maketitle

\thispagestyle{fancy}


\section{Introduction}

The past 20 years have seen unprecedented growth in the field of gamma-ray astronomy. Highly successful missions such as the Compton Gamma-Ray Observatory (CGRO) detected gamma-ray sources and a more fundamental understanding of the basic physical processes involved for those sources. From 1991 to 1995, the Burst and Transient Source Experiment (BATSE) on CGRO, which was sensitive from 23 keV to 1.8 MeV, was the only instrument which was capable of monitoring the low energy gamma-ray sky. In 1995, the Rossi X-Ray Timing Explorer (RXTE) satellite was launched, but it had a relatively narrow field of view and was only sensitive up to $\sim 250$ keV. Gamma-ray Burst Monitor (GBM) currently flying on the Fermi spacecraft since 2008 is the only instrument performing all-sky observations above $\sim 200$ keV and below 20 MeV at present.

Therefore, the combination of BATSE and GBM all-sky analysis gives us a $\sim 20$ year history of the hard X-ray/soft gamma-ray sky.

\section{Earth Occultation Technique}

As demonstrated with BATSE \cite{Harmon2002}\cite{Ling2000}  and GBM \cite{Wilson2012}, monitoring known hard x-ray/soft gamma-ray sources is possible with simple non-pixellated or low spatial resolution detectors by using the Earth as a temporal modulator of the gamma-ray flux. This is the Earth Occultation Technique (EOT). The Earth, seen from a BATSE detector, has a diameter of $\sim 140^{\circ}$. When the source sets below the Earth's limb, a downward step appears in the detector count rate (and vice-versa for rising steps). The height above the background of the steps gives a direct measurement of the count rate of the occulted source.

Occultation times can be predicted using the coordinates of the source and the spacecraft position. The occultation time is defined to be the time when the transmission through the atmosphere is $50\%$ at 100 keV. 

The Enhanced BATSE Occultation Package (EBOP) was developed by the Jet Propulsion Laboratory (JPL) group, and was applied to BATSE data to search and fit the occultation steps \cite{Ling2000}.
For known sources, a fitting procedure can be used on a 240 second fitting window centered at the expected occultation time bin. For each detector and energy channel the model fit consists of a quadratic curve describing the background and a set of response vectors that model the effects of all sources contained within the fit window. However, the sensitivity of the step searching technique is limited by the difficulty in accurately determining the background over several orbits and by non-statistical noise introduced by bright pulsating sources with periods of a few minutes, such as Vela X-1 and the unknown sources. Hence, unknown sources affect the EBOP results systematically.

\section{Earth Occultation Imaging Technique with Differential Filter}

\begin{figure}[!hbp]
\begin{center}
\includegraphics [width=80mm]{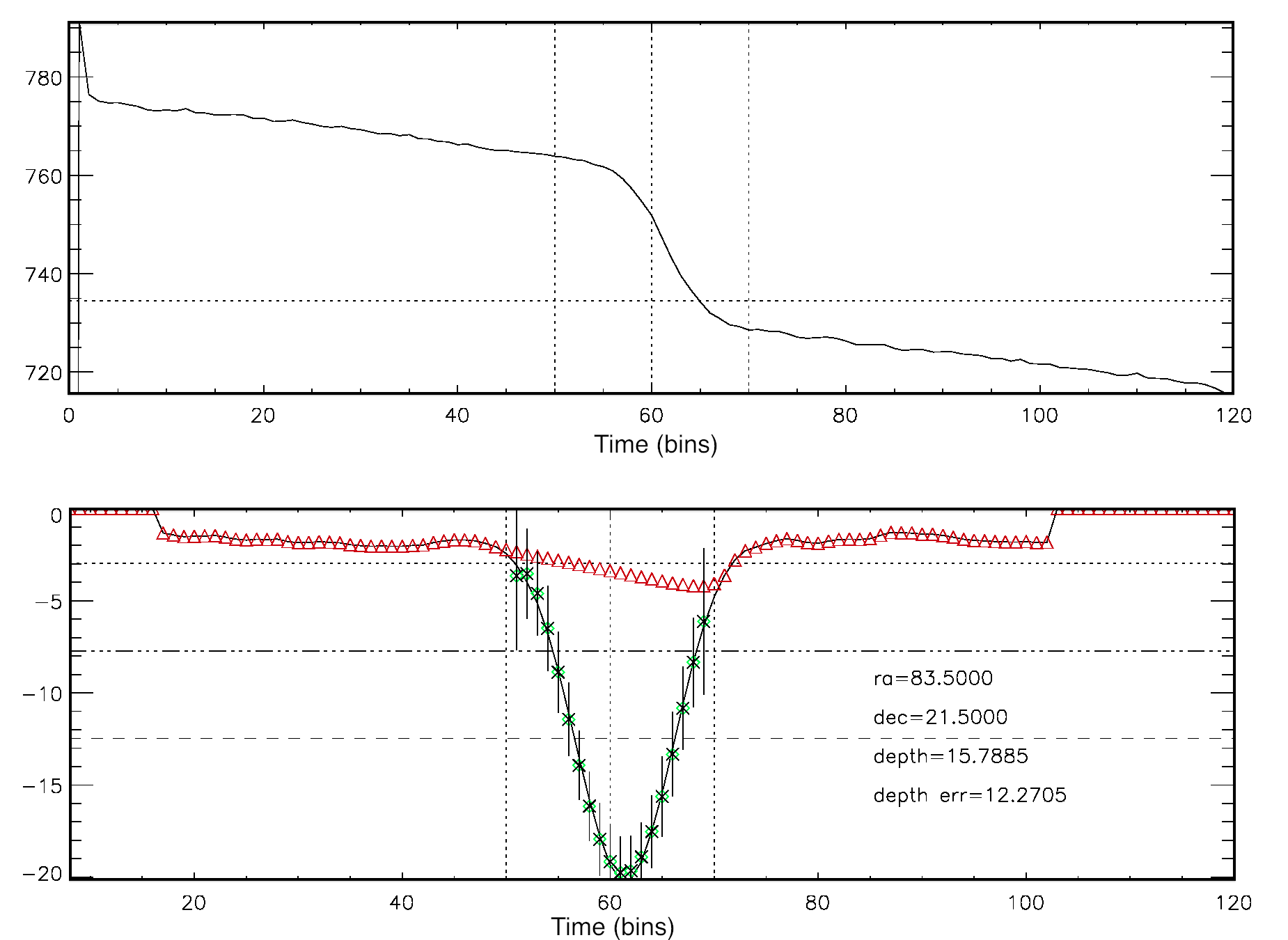}
\caption{\emph{Top}: Example of an occultation window for the Crab.  \emph{Bottom}: The filtered data with the central portion fit to a polynomial (green), and the outer background portion fit to a spline function (red). }
\label{fig:occ_fit}
\end{center}
\end{figure}

Since EOT and EBOP require an input catalog with predetermined source position, another method is needed to search for unknown sources. An imaging method conceptually similar to x-ray computed tomography used in medical imaging has been developed \cite{Rodi2011}.

When a source occults, the Earth's limb can be projected onto the sky over which counts are integrated. During the orbital precession of the spacecraft, the angle between the source position and the plane of the satellite orbit changes. Thus over an orbital procession period a range of projections are sampled. The projections from rising and setting occultation steps can be combined to generate an image of the sky and localize a source. 

The angular resolution of this method is limited by the finite time required to go from full transmission to $50\%$ attenuation for a source. 
For a typical orbital speed of $8 km/s$, the duration of an occultation step is \( \sim 8\) seconds for normal incidence.  As the angle between the orbital plane and the source increases, the duration of the occultation step increases by \( \sim 1/ \cos( \beta ) \), where \( \beta \) is the elevation angle between the source and the orbital plane of the satellite.  The angular resolution is \( 360^{\circ} \times \Delta t / P \), where \( \Delta t \) is the occultation duration and \( P \) is the satellite's orbital period ( \( \sim 90 \) minutes).  The angular resolution ranges from \( \sim 0.5^{\circ} \text{ at } \beta = 0^{\circ} \text{ to } \sim 1.25^{\circ} \text{ at } \beta = 66^{\circ} \).  When \( \beta > 66^{ \circ} \), occultation no longer occur.

Instead of a catalog of known source positions, our imaging technique starts with a catalog of virtual sources covering the whole sky with spacing of $\sim 0.25^{\circ}$, resulting in a catalog of $\sim 660,000$ positions on the sky. With these virtual source positions and the spacecraft position history, occultation times can be predicted. To be able to combine rise and set steps together, occultation windows for rise steps are converted to set steps by rotating the data about the occultation time at the centre of the 4-min window. A weighted average of all the windows for a virtual source is calculated for an orbital procession period. Then, the averaged window is passed through a differential filter of the form \cite{Shaw2004},
\begin{equation}\label{eq:df}
o_i = \frac{ \sum_{j=i+f_a}^{j=i+f_a+f_b} r_j - \sum_{j=i-f_a}^{j=i-f_a-f_b} r_j}{f_b},
\end{equation}

where \( r_j \) is the number of counts in bin \( j \), \( f_a \) is the inner boundary of the filter, and \( f_b \) is the outer boundary (citation needed).  The values of \( f_a \text{ and } f_b \) used here are 3 bins and 8 bins, respectively.  The differential filter is an approach to transform those step features into gaussian-like peaks, where the amplitude of the peak is related to the intensity of the source.  To calculate the amplitude, the window is fit in two sections. The central part, within \( \pm (2 f_a + f_b ) \) bins from the middle, is fit to quadratic, cubic, and quartic function, and the fit giving the lowest reduced $\chi^{2}$ will be kept. The outer portions of the window are fit with a spline function and joined by a straight line crossing the central region.  The amplitude of the virtual source is determined by taking the difference between the two fits at the occultation time.

Fig.~\ref{fig:occ_fit} shows an example of an occultation window and the result of filtering for a virtual source very near the Crab. In fig.~\ref{fig:occ_fit}, the middle dotted vertical line corresponds to the expected occultation time of the virtual source.

\section{Data Selection and Source Confusion}

\begin{figure*}[t]
\begin{center}
\includegraphics [ width=0.85\textwidth]{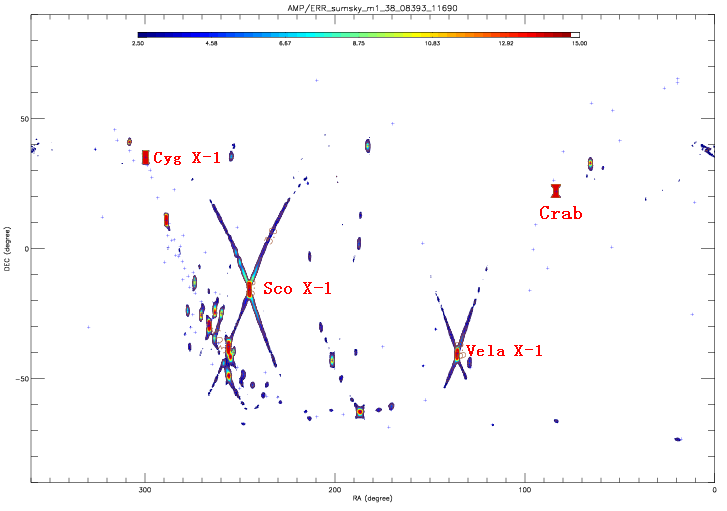}
\caption{Example of Source Confusion ''Arms''. Source confusion algorithm has been applied to Crab and Cyg X-1, but not Sco X-1 nor Vela X-1. Obvious pattern differences are shown in the map.}
\label{fig:arm}
\end{center}
\end{figure*}

Prior to application of the Earth occultation method, proper data selection is required to remove large fluctuations that may affect the imaging. This selection process contains three steps.
The first step is performed by using Quality Flags. Any high-rate events, which triggered the GRB acquisition mode, are flagged. 
The second step is due to South Atlantic Analomy (SAA) passage, where data bins within 20 seconds before and after SAA passage are excluded.
The last step consists of additional flagging of very short cosmic-ray events, which normally occur in only one detector at a time. 

During the Earth occultation imaging process, another systematic effect is source confusion caused by bright sources. The issue of source confusion arises from the location ambiguity of where along the Earth's limb the measured amplitude originates. Since BATSE detectors do not have imaging capability, the flux for an occultation time could be allocated to anywhere along the Earth's limb. Hence, bright sources cause ''X'' patterns that can extend for tens of degrees. These ''arms'' make it difficult to search for fainter sources along the ''arms''. To minimize the effects of source confusion, an algorithm has been developed to ignore occultation windows for virtual sources when bright sources occult close in time (\( < 11 \) seconds). Fig.~\ref{fig:arm} illustrates the result of source confusion. The algorithm is applied here to the Crab and Cyg X-1. Hence, no bright arms originated from these two bright sources. But Vela X-1 and Sco X-1 are associated with the strong ''X'' patterns, since the algorithm was not applied here to them.

\section{Results}

\begin{figure*}[t]
\centering
\includegraphics[angle=90,width=0.9\textwidth]{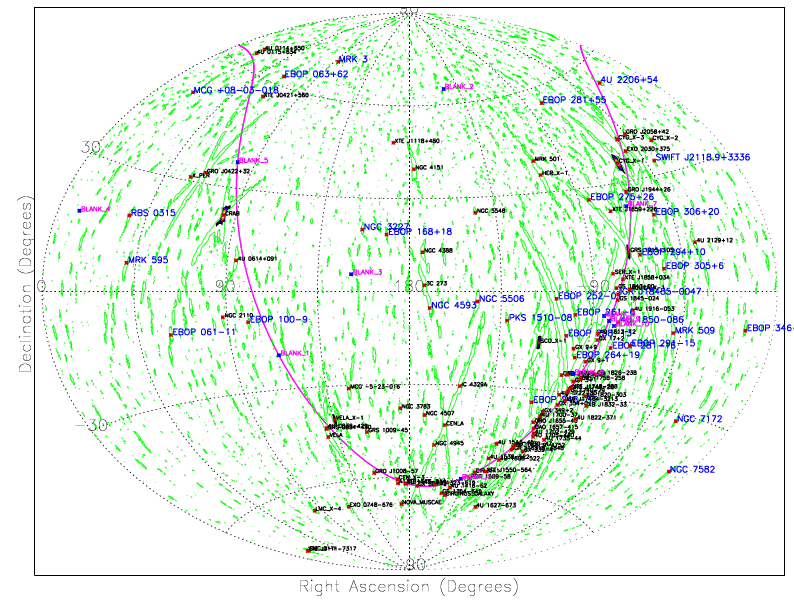}
\vspace{0in}\hspace{6in}\parbox{6.5in}{\caption{All-sky map summed over 64 precession periods (TJD 08393 - TJD 11690) over the energy range 23 - 98 keV. 3 sigma (statistical) features are contoured in green. he source names in black denote the original sources in the 2007 catalog; blue denotes the 32 new sources; magenta marks 11 blank sources included to study systematic effects and sensitivity}}
\label{fig:allskymap}
\end{figure*}

The imaging algorithm has been applied to 64 precession periods, 9 years of data, from 1991 to 2000, to generate all-sky images at 23-98 keV. Fig.~\ref{fig:allskymap} is a significance map with green contours at 3 $\sigma$. In Fig.~\ref{fig:allskymap}, the source names in black denote the original sources in the 2007 catalog; blue denotes the 32 new sources; magenta marks 11 blank sources included to study systematic effects and sensitivity. These sources were identified through cross-correlating features in the maps with sources in the Swift/BAT, INTEGRAL/SPI, and Fermi/GBM catalogs.

126 sources are included in the new BATSE/EBOP 2012 catalog. Of those, 94 sources coincide with sources in the 2007 EBOP catalog, and 32 are new sources, with 15 of them identified with known Swift and INTEGRAL sources, and 17 unidentified. Table.~\ref{table:new_sources} shows the name and position of the new identified sources.

\begin{table}[t]
\begin{center}
\caption{15 New known Sources found with Earth Occultation Imaging.}
\begin{tabular}{|l|c|c|}
\hline \textbf{Source Name} & \textbf{RA} & \textbf{Dec}
\\
\hline	MCG+08-03-018	&	20.67	&	50.07	\\
\hline	RBS0315	&	36.26	&	18.81	\\
\hline	MRK595	&	40.46	&	7.19	\\
\hline	MRK3	&	93.94	&	71.03	\\
\hline	NGC3227	&	155.87	&	19.86	\\
\hline	NGC4593	&	189.91	&	-5.35	\\
\hline	NGC5506	&	213.31	&	-3.21	\\
\hline	PKS1510-08	&	228.22	&	-9.09	\\
\hline	IGRJ18485-0047	&	282.17	&	-0.79	\\
\hline	4U1850-086	&	283.27	&	-8.71	\\
\hline	MRK509	&	311.03	&	-10.73	\\
\hline	SWIFTJ2118.9+3336	&	319.86	&	33.56	\\
\hline	NGC7172	&	330.51	&	-31.87	\\
\hline	4U2206+54	&	331.96	&	54.51	\\
\hline	NGC7582	&	349.60	&	-42.37	\\

\hline
\end{tabular}
\label{table:new_sources}
\end{center}
\vspace{-10mm}
\end{table}

\begin{table}[t]
\begin{center}
\caption{17 Unidentified BATSE Sources found with Earth Occultation Imaging.}
\begin{tabular}{|l|c|c|}
\hline	\textbf{Unidentified Source}	&	\textbf{RA}	&	\textbf{Dec}	\\
\hline	EBOP061-11	&	61.19	&	-11.70	\\
\hline	EBOP063+62	&	63.37	&	62.55	\\
\hline	EBOP100-9	&	100.24	&	-9.17	\\
\hline	EBOP168+18	&	168.40	&	18.40	\\
\hline	EBOP248-33	&	248.39	&	-33.92	\\
\hline	EBOP252-01	&	252.02	&	-2.19	\\
\hline	EBOP258-13	&	258.30	&	-13.26	\\
\hline	EBOP261-6	&	261.63	&	-6.93	\\
\hline	EBOP264-19	&	264.47	&	-19.53	\\
\hline	EBOP275+26	&	275.37	&	26.49	\\
\hline	EBOP281+55	&	281.38	&	55.51	\\
\hline	EBOP281-16	&	281.41	&	-16.05	\\
\hline	EBOP291-15	&	290.82	&	-15.01	\\
\hline	EBOP294+10	&	294.25	&	10.01	\\
\hline	EBOP305+6	&	305.23	&	6.00	\\
\hline	EBOP306+20	&	306.03	&	20.29	\\
\hline	EBOP346-8	&	346.18	&	-8.69	\\

\hline
\end{tabular}
\label{table:unidentified_sources}
\end{center}
\vspace{-10mm}
\end{table}

Comparison with the current GBM catalog \cite{Wilson2012}, it shows 69 of the BATSE sources coinciding with the GBM catalog. Of these, 3 are new sources found by the current imaging analysis, NGC5508, PKS 1510-08, and 4U 2206+54.

\bigskip 


\begin{acknowledgments}
This material is based upon work supported by the Louisiana Optical Network Institute (LONI).  Portions of this research were conducted with high performance computational resources provided by Louisiana State University (http://www.hpc.lsu.edu).  This work is supported by NASA/Louisiana Board of Regents Cooperative Agreement NNX07AT62A.

The work described in this paper was carried out at the Jet Propulsion Laboratory under the contract with the National Aeronautics and Space Administration.

\end{acknowledgments}

\bigskip 

\end{document}